\newcommand{\pp} {\mbox{$p+p$}}
\newcommand{\pt} {\mbox{$p_T$}}
\newcommand{\piz} {\mbox{$\pi^0$}}
\newcommand{\pbar} {\mbox{$\overline{p}$}}
\newcommand{\snn} {\mbox{$\sqrt{s_{NN}}$}}
\newcommand{\npart} {\mbox{$N_{\it part}$}}
\newcommand{\ncoll} {\mbox{$N_{\it coll}$}}

\documentclass[
    ,final            
  ]
  {aipproc}

\layoutstyle{6x9}

\begin{document}
\title{Particle Composition at High $p_{T}$ in Au+Au Collisions at $\sqrt{s_{NN}}$ = 200 GeV}
\author{Tatsuya Chujo}
{
  address={Brookhaven National Laboratory, Upton, NY 11973-5000, USA}
}

\begin{abstract}
We report the recent results of proton and anti-proton yields 
as a function of centrality and $p_T$ in Au+Au collisions at 
$\sqrt{s_{NN}}$ = 200 GeV, measured by the PHENIX experiment at RHIC. 
In central collisions at intermediate transverse momenta 
($1.5 < \pt < 4.5$\,GeV/$c$) a significant fraction of all produced 
particles is protons and anti-protons. They show a different scaling
behavior from that of pions.  The $\pbar/\pi$ and $p/\pi$ ratios are 
enhanced compared to peripheral Au+Au, p+p and $e^{+}e^{-}$ collisions. 
This enhancement is limited to $\pt < 5$\,GeV/$c$ as deduced from the 
ratio of charged hadrons to $\piz$ measured in the range 
$1.5 < \pt < 9$\,GeV/$c$.
\end{abstract}

\maketitle

Heavy-ion collisions at RHIC energies allow us to study the properties 
of nuclear matter at extreme energy densities. High $\pt$ hadrons production
originating in the fragmentation of partons with a large momentum transfer 
(hard processes) are sensitive probes of the hottest and densest stage of 
the collision. One of the most significant results from the first year of 
RHIC run was the suppression of yields both for charged and $\piz$ at high $\pt$ 
in central Au+Au with respect to the number of nucleon-nucleon collisions 
($\ncoll$)~\cite{ppg003,ppg013}. Moreover, it was found that $\piz$ yields 
are more strongly suppressed than for charged hadrons~\cite{ppg003}, and 
the yields of $p$ and $\pbar$ near 2\,GeV/$c$ in central collisions
are comparable to the yield of pions~\cite{ppg006}. These observations 
suggest that a significant fraction of all particle yields is $p$ and 
$\pbar$ at the intermediate $\pt$ in central Au+Au collisions. 
We present here the results of $p$ and $\pbar$ yields including their scaling 
properties and ratios of $p/\pi$, $\pbar/\pi$ as a function of centrality
in Au$+$Au collisions at $\snn= 200$ GeV measured by the PHENIX experiment~\cite{NIM}. 
The detailed analysis methods and results are found in 
references~\cite{ppg015,ppg026} for identified charged hadrons, in 
reference~\cite{ppg014} for $\piz$, and in reference~\cite{ppg023} for 
inclusive charged hadrons. 

\begin{figure}[t]
\caption{$p/\pi$ (left) and $\pbar/\pi$ ratios for central(0-10\%), 
mid-central(20-30\%) and peripheral (60-92\%) Au+Au collisions at 
$\snn = 200$ GeV~\cite{ppg015,ppg026}. Open (filled) points are 
for charged (neutral) pions, respectively. Data from $\sqrt{s} = 53 $ 
GeV $\pp$ collisions~\cite{ISR} are shown with stars. The dashed and 
dotted lines are $(\pbar+p)/(\pi^{+}+\pi^{-})$ ratio in gluon and 
in quark jets~\cite{DELPHI}.}
\includegraphics[scale=0.7]{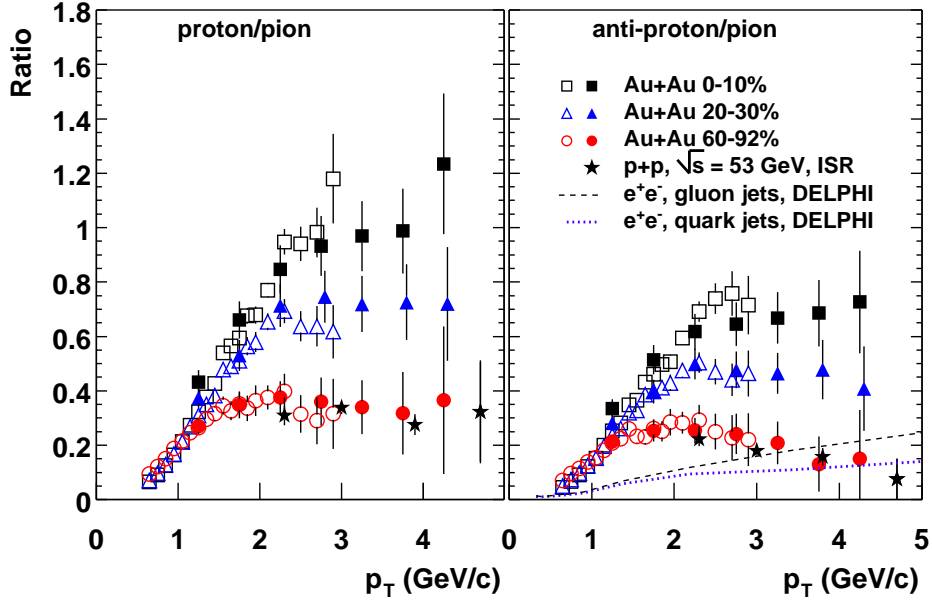}
\label{fig:1}
\end{figure}

\begin{figure}[t]
\caption{Nuclear modification factor $R_{CP}$ for $(p+\pbar)/2$ 
(filled circles) and $\piz$~\cite{ppg015,ppg026}. Dashed and dotted 
lines indicate $\ncoll$ and $\npart$ (number of participant nucleons) 
scaling; the shaded bars show the systematic errors on these quantities.}
\includegraphics[scale=0.5]{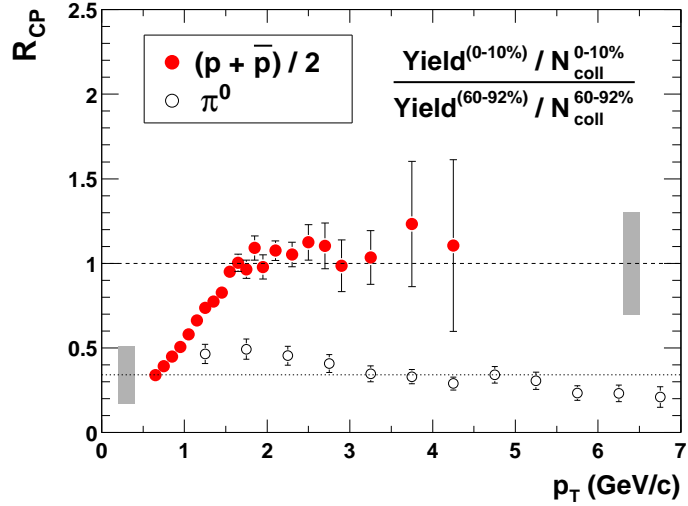}
\label{fig:2}
\end{figure}

Figure~\ref{fig:1} shows the $p/\pi$ and $\pbar/\pi$ ratios as a
function of $\pt$ measured at mid-rapidity in central (0--10\%),
mid-central (20--30\%), and peripheral (60--92\%) Au$+$Au collisions
at $\snn = 200$ GeV. For all centralities the ratios rise steeply at 
low $\pt$ and then, at a value of $\pt$ which increases from peripheral 
to central collisions, level off. In central collisions the ratios are 
a factor of $\sim 3 $ larger than in peripheral events. At $\pt > 2$\,GeV/$c$ 
the peripheral Au$+$Au data agree well with the ratios observed 
in $\pp$ collisions at lower energies~\cite{ISR}. Above 3 GeV/$c$ 
the $p/\pi$, $\pbar/\pi$ ratios in peripheral collisions are also 
consistent with gluon and quark jet fragmentation~\cite{DELPHI}. Deviations from 
jet fragmentation below 3 GeV/c indicate the absence of soft hadron 
production in the $e^{+}e^{-}$ data. In Figure~\ref{fig:2}, we compare 
the $\ncoll$ scaled central to peripheral yield ratios, $R_{\rm CP}$, 
for $(p+\pbar)/2$ and $\piz$. In the $\pt$ range from 1.5 to 4.5 GeV/$c$, 
$p$ and $\pbar$ are not suppressed in contrast to $\piz$ which are largely suppressed 
by a factor of 2-3. Moreover, this behavior holds for all centralities 
(see references~\cite{ppg015,ppg026}), while the suppression in the 
$\piz$ yields increases from peripheral to central collisions~\cite{ppg014}. 

It is interesting why the suppression for $p$ and $\pbar$ is absent in central Au+Au
collisions. Recently the observed abundance of protons yields relative to 
pions in central collisions has been attributed to the recombination of quarks, 
rather than fragmentation~\cite{recombination}. In this model, recombination 
for $p$ and $\pbar$ is effective up to $\pt \simeq 5$\,GeV above which 
fragmentation dominates for all particle species. Another explanation of 
the observed large baryon content invokes a topological gluon configuration: 
the baryon junction~\cite{bjunctions}. A centrality dependence, which is in 
qualitative agreement with the results presented here, has been 
predicted~\cite{ptopi_bjunctions}. In both theoretical models, the baryon/meson 
enhancement is limited to $\pt < $ 5--6 GeV/$c$. In order to test these
theoretical predictions, we measure charged hadrons to $\piz$ measured in
$1.5 < \pt < 9$\,GeV/$c$ (see references~\cite{ppg015,ppg023}). 
It is found that in central collisions for $1 < \pt<$ 4.5 GeV/$c$, $h/\piz$ 
ratio is enhanced by as much as 50\% above the $\pp$ value. Above 
$\pt \simeq 5$\,GeV/$c$, the particle composition is consistent with that 
measured in $\pp$ collisions. This indicates that the scaling of the proton 
yields should become consistent with that of pions at $\pt > 5$~GeV/c. 
Similar limiting behavior of baryon/meson enhancement is observed in 
$\Lambda$ and $K^{0}_{S}$ by the STAR collaboration~\cite{star_k0}. 
It is possible that nuclear effects such as the ``Cronin effect''~\cite{cronin} 
contribute to the observed large (anti)proton/pion ratios. The recent results of 
inclusive charged hadrons and $\piz$ in d$+$Au at $\snn=200$\,GeV suggest that 
the Cronin effect in baryons is different from that in mesons~\cite{ppg028}. 
Detailed studies of particle composition in d$+$Au collisions will help our 
understanding of the baryon production at the intermediate $\pt$ region at RHIC.

{\footnotesize
We thank the staff of the Collider-Accelerator and Physics
Departments at BNL for their vital contributions.  We acknowledge
support from the Department of Energy and NSF (U.S.A.), MEXT and
JSPS (Japan), CNPq and FAPESP (Brazil), NSFC (China), CNRS-IN2P3
and CEA (France), BMBF, DAAD, and AvH (Germany), OTKA (Hungary), 
DAE and DST (India), ISF (Israel), KRF and CHEP (Korea),
RMIST, RAS, and RMAE, (Russia), VR and KAW (Sweden), U.S. CRDF 
for the FSU, US-Hungarian NSF-OTKA-MTA, and US-Israel BSF.
}
\bibliographystyle{aipproc}
\bibliography{rhi-fri-1-6-chujo}

\hyphenation{Post-Script Sprin-ger}
\begin{thebibliography}{1}
\expandafter\ifx\csname natexlab\endcsname\relax\def\natexlab#1{#1}\fi
\providecommand{\enquote}[1]{``#1''}
\expandafter\ifx\csname url\endcsname\relax
  \def\url#1{\texttt{#1}}\fi
\expandafter\ifx\csname urlprefix\endcsname\relax\def\urlprefix{URL }\fi

\bibitem{ppg003} PHENIX Collaboration, K.~Adcox {\it et al.},
 \Journal{\PRL} {88}{022301}{2002}.
\bibitem{ppg013} PHENIX Collaboration, K.~Adcox {\it et al.},
 \Journal{\PLB} {561}{82}{2003}.
\bibitem{ppg006} PHENIX Collaboration, K.~Adcox {\it et al.},
 \Journal{\PRL} {88}{242301}{2002}.
\bibitem{NIM} PHENIX Collaboration, K.~Adcox {\it et al.},
 \Journal{\NIMA}{499}{469-479}{2003}.
\bibitem{ppg015} PHENIX Collaboration, S.~S.~Adler~{\it et al.},
 submitted to \PRL, nucl-ex/0305036.
\bibitem{ppg026} PHENIX Collaboration, S.~S.~Adler~{\it et al.},
 submitted to \PRC, nucl-ex/0307022.
\bibitem{ppg014} PHENIX Collaboration, S.~S.~Adler~{\it et al.},
 \Journal{\PRL} {91}{072301}{2003}, [nucl-ex/0304022].
\bibitem{ppg023} PHENIX Collaboration, S.~S.~Adler~{\it et al.},
 submitted to \PRC, nucl-ex/0308006.
\bibitem{ISR} B.~Alper {\it et al.},
 \Journal{\NPB}{100}{237-290}{1975}.
\bibitem{DELPHI} DELPHI Collaboration, P.~Abreu {\it et al.}, 
 \Journal{EPJC}{17}{207}{2000}.
\bibitem{recombination}
 R.~C.~Hwa~{\it et al.}, Phys.\ Rev.\ C {\bf 67}, 034902 (2003);
 R.~J.~Fries~{\it et al.}, nucl-th/0301087;
 V.~Greco~{\it et al.}, nucl-th/0301093.
\bibitem{bjunctions}
 G.C.~Rossi~{\it et al.}, \Journal{\NPB}{123}{507}{1977};
 D.~Kharzeev, \Journal{\PLB} {378} {238} {1996};
 S.E.~Vance~{\it et al.}, \Journal{\PLB}{443}{45}{1998}.
\bibitem{ptopi_bjunctions} 
 I.~Vitev~{\it et al.}, \Journal{\PRC}{65}{041902}{2002};
 I.~Vitev~{\it et al.}, hep-ph/0109198.
\bibitem{star_k0} STAR Collaboration, J.~Adams~{\it et al.},
 submitted to \PRL, nucl-ex/0306007.
\bibitem{cronin}
 J.~Cronin~{\it et al.}, \Journal{\PRD}{11}{3105}{1975}.
 D.~Antreasyan~{\it et al.}, \Journal{\PRD}{19}{764}{1979}.
\bibitem{ppg028} PHENIX Collaboration, S.~S.~Adler~{\it et al.},
 submitted to \PRL, nucl-ex/0306021.

\end{thebibliography}

\IfFileExists{\jobname.bbl}{}
 {\typeout{}
  \typeout{******************************************}
  \typeout{** Please run "bibtex \jobname" to obtain}
  \typeout{** the bibliography and then re-run LaTeX}
  \typeout{** twice to fix the references!}
  \typeout{******************************************}
  \typeout{}
 }

\end{document}